\documentclass[fleqn,twoside]{article}
\usepackage{espcrc2}
\usepackage{latexsym}
\usepackage{euscript}
\usepackage{booktabs}

\usepackage{graphicx}
\usepackage[figuresright]{rotating}

\def\simge{\mathrel{%
   \rlap{\raise 0.511ex \hbox{$>$}}{\lower 0.511ex \hbox{$\sim$}}}}   
\def\simle{\mathrel{   
   \rlap{\raise 0.511ex \hbox{$<$}}{\lower 0.511ex \hbox{$\sim$}}}}   
\def\slashchar#1{\setbox0=hbox{$#1$}           
   \dimen0=\wd0                                 
   \setbox1=\hbox{/} \dimen1=\wd1               
   \ifdim\dimen0>\dimen1                        
      \rlap{\hbox to \dimen0{\hfil/\hfil}}      
      #1                                        
   \else                                        
      \rlap{\hbox to \dimen1{\hfil$#1$\hfil}}   
      /                                         
   \fi}                                         %

\def\simge{\mathrel{%
   \rlap{\raise 0.511ex \hbox{$>$}}{\lower 0.511ex \hbox{$\sim$}}}}   
\def\simle{\mathrel{   
   \rlap{\raise 0.511ex \hbox{$<$}}{\lower 0.511ex \hbox{$\sim$}}}}   
\def\slashchar#1{\setbox0=\hbox{$#1$}           
   \dimen0=\wd0                                 
   \setbox1=\hbox{/} \dimen1=\wd1               
   \ifdim\dimen0>\dimen1                        
      \rlap{\hbox to \dimen0{\hfil/\hfil}}      
      #1                                        
   \else                                        
      \rlap{\hbox to \dimen1{\hfil$#1$\hfil}}   
      /                                         
   \fi}

\newcommand\mycaption[1]{\caption{\footnotesize #1}}

\newcommand{\AmS}{{\protect\the\textfont2
  A\kern-.1667em\lower.5ex\hbox{M}\kern-.125emS}}

\newcommand{\ba}{\begin{equation} \left\{ \begin{array}{lr}}
\newcommand{\ea}{\end{array} \right. \end{equation}}

\newcommand{\bea}{\begin{eqnarray}}
\newcommand{\eea}{\end{eqnarray}}
\newcommand{\beqa}{\begin{eqnarray}}
\newcommand{\eeqa}{\end{eqnarray}}
\newcommand{\be}{\begin{equation}}
\newcommand{\ee}{\end{equation}}
\newcommand{\beq}{\begin{equation}}
\newcommand{\eeq}{\end{equation}}

\newcommand{\Tr}{\mbox{Tr}\;}

\title{Remarks on the discretization of physical momenta in lattice QCD\thanks{Talk given at Lattice 2004}}

\author{Nazario~Tantalo\address[ROME2]{Dipartimento di Fisica, Universit\`a di Roma ``Tor Vergata'', 
        V. R. Scientifica 1, I-00133 Rome, Italy}\address[INFN]{INFN Roma 2, V. R. Scientifica 1, I-00133 Rome, Italy}}
       
\begin{document}

\begin{abstract}
The calculation on the lattice of cross--sections, form--factors and decay rates 
associated to phenomenologically relevant physical processes is complicated by the spatial momenta
quantization rule arising from the introduction of limited box sizes in numerical simulations.
A method to overcome this problem, based on the adoption
of two distinct boundary conditions for two fermions species on a finite 
lattice, is here discussed and numerical results supporting
the physical significance of this procedure are shown.
\end{abstract}

\maketitle

\section{The problem}
\label{sec:problem} 

In the formulation of quantum field theories on a lattice 
the introduction of a finite volume is unavoidable 
when numerical simulations are used as tool of investigation.
As a consequence of the limited box size, spatial momenta
come out to be quantized accordingly to the choice of the
boundary conditions.

The momentum quantization represents a severe limitation in various 
phenomenological applications by making impossible, in the great
majority of the cases, to calculate
cross--sections, decay--rates and form--factors
in the interesting kinematical region or at the physical values
of the masses of the particles involved in the processes.
In this talk it is discussed a method, previously introduced
in ref.~\cite{deDivitiis:2004kq}, that allows to overcome these
difficulties.
The idea consists in using the dependence of the momentum
quantization condition upon the choice of the boundary conditions (BC)
and in requiring different fermion species to satisfy
different BC.

\section{The way out}
\label{sec:method}

In order to clarify the dependence of the  
momentum quantization rule upon the choice of the 
boundary conditions let us first consider
the case of a particle satisfying periodic boundary conditions (PBC).
For a fermionic field $\psi(x)$ 
on a 4--dimensional finite volume of topology $T\times L^3$
with PBC in the spatial directions one has
\beq
\psi(x+\vec{e}_i\ L) = \psi(x)\;, \qquad i=1,2,3
\eeq
This condition can be re-expressed by Fourier transforming
both members of the previous equation
\beq
\int{d^4p\ e^{-ip(x+\vec{e}_i\ L)}\ \tilde{\psi}(p)}
= \int{d^4p\ e^{-ipx}\ \tilde{\psi}(p)}
\label{eq:fspbc}
\eeq
and implies 
\beq
e^{ip_iL} = 1 \quad \Longrightarrow \quad p_i = \frac{2\pi\ n_i}{L}
\label{eq:pbcmom}
\eeq
where the $n_i$'s are integer numbers.
The authors 
of~\cite{Jansen:1996ck,Bucarelli:1998mu,Guagnelli:2003hw,Bedaque:2004kc,Gross:1982at,Kiskis:2002gr,Kiskis:2003rd,Roberge:1986mm} 
have  considered a generalized set of boundary conditions, that here
we call $\theta$--boundary conditions ($\theta$--BC),
depending upon the choice of a topological 3--vector $\vec{\theta}$
\beq
\psi(x+\vec{e}_i\ L) = e^{i\theta_i}\ \psi(x)\;, \qquad i=1,2,3
\label{eq:thetabc}
\eeq
The modification of the boundary conditions affects the zero of the 
momentum quantization rule. 
Indeed, by re-expressing equation~(\ref{eq:thetabc}) in Fourier space, 
as already done in the case of PBC in equation~(\ref{eq:fspbc}), one
has
\beq
e^{i(p_i-\frac{\theta_i}{L})L} = 1 \quad \Longrightarrow 
\quad p_i = \frac{\theta_i}{L} + \frac{2\pi\ n_i}{L}
\label{eq:thetamom}
\eeq
It comes out that the spatial momenta are still quantized as for
PBC but shifted by an arbitrary \emph{continuous} 
amount ($\theta_i/L$).
The physical significance of this continuous shift in the
allowed momenta becomes manifest, from the theoretical
point of view, by realizing that
eq.~(\ref{eq:thetabc}) is the condition satisfied
by the electron wavefunctions in a periodic
solid crystal accordingly to the well known Bloch's theorem.
Indeed, quarks on a lattice experience a periodic potential
(the gauge fields) exactly as the electrons do in a crystal and
the results of the Bloch's theorem apply straight forwardly
also to the lattice fields.

The generalized $\theta$--dependent
boundary conditions of equation~(\ref{eq:thetabc}) can be  implemented by
making a unitary Abelian transformation on the fields satisfying $\theta$--BC
\beq
\psi(x) \quad \longrightarrow  \quad 
{\mathcal U}\ (\theta,x) \psi(x)= e^{-\frac{i \theta x}{L}}\  \psi(x)
\label{eq:unittransf}
\eeq
As a consequence of this transformation the resulting field satisfies
periodic boundary conditions but obeys a modified Dirac equation
\beqa
&S[\bar{\psi},\psi]& \rightarrow
\nonumber \\
& \rightarrow&   \sum_{x,y}{ \bar{\psi}(x)\ {\mathcal U}(\theta,x)D(x,y) {\mathcal U}^{-1}(\theta,y)\ \psi(y)}
\nonumber \\
& =&  \sum_{x,y}{ \bar{\psi}(x) \ D_\theta(x,y)\ \psi(y)} 
\label{eq:Smod}
\eeqa    
where the $\theta$--dependent lattice Dirac operator $D_\theta(x,y)$ is obtained
by starting from the preferred discretization of the Dirac operator and 
by modifying the definition of the covariant lattice derivatives
(see ref.~\cite{deDivitiis:2004kq} for details).

\section{Numerical tests}

%
%
%
%
%
\begin{figure}[t]
\begin{center}
\includegraphics[width=\columnwidth]{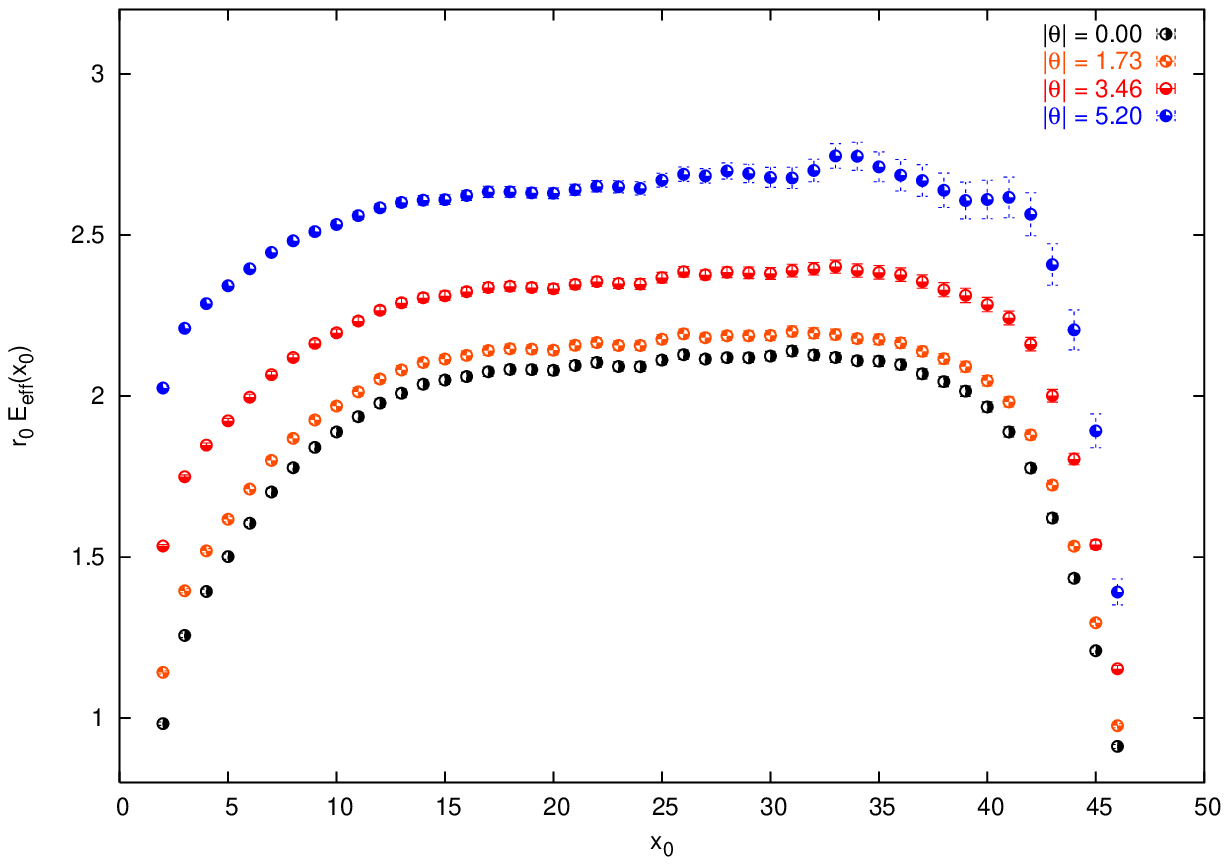}
\vskip -20pt
\mycaption{Effective energies $E_{eff}^{ij}(\theta,a;x_0)$
at fixed cut--off. The results correspond to the simulation done at $\beta=6.211$
with $r_0\ m_1^{RGI} = 0.655$ and $r_0\ m_2^{RGI} = 0.354$.
Similar figures could have been shown for other combinations of the simulated quark masses
and for the other values of the bare coupling.}
\label{fig:effmass}
\end{center}
\vskip -20pt
\end{figure}

In order to prove numerically that the term $\vec{\theta}/L$ acts as a true
physical momentum, one can study  the energy of a meson
made up by two different quarks with different  $\theta$--BC for the two 
flavors.
In the following 
we work in the $O(a)$--improved Wilson--Dirac lattice formulation of the 
QCD within the Schr\"odinger Functional formalism but,
we want to stress that the use of $\theta$--BC in the spatial directions 
is completely decoupled from the choice
of time boundary conditions and 
can be profitably used outside the Schr\"odinger Functional formalism,
for example in the case of standard periodic time boundary conditions. 
Let us consider the following correlators
\beqa
&&f_P^{ij}(\theta;x_0) = 
\nonumber \\
&&=-\frac{a^6}{2}\sum_{\vec{y},\vec{z},\vec{x}}{
\langle\ \bar{\zeta}_i(\vec{y})\gamma_5 \zeta_j(\vec{z}) \ 
\bar{\psi}_j(x) \gamma_5 \psi_i(x) \ \rangle
} 
\label{eq:fp}
\eeqa
where $i$ and $j$ are flavor indices, all the fields satisfy periodic boundary conditions 
and the two flavors obey different $\theta$--modified Dirac equations, as explained in eq.~(\ref{eq:Smod}). 
In practice it is adequate to choose the flavor $i$
with $\theta=0$, i.e. with ordinary PBC, and the flavor $j$ with $\theta\neq 0$.
After the Wick contractions the pseudoscalar correlator of
equation~(\ref{eq:fp}) reads
\beqa
&&f_P^{ij}(\theta;x_0) = 
\nonumber \\
&&=\frac{a^6}{2}\sum_{\vec{y},\vec{z},\vec{x}} \Tr{
\langle \gamma_5 S_j(\theta;\vec{z},x) \gamma_5 S_i(0;x,\vec{y}) \rangle
} 
\label{eq:fpcontractions}
\eeqa
where $S(\theta;x,y)$ and $S(0;x,y)$ are the inverse of the $\theta$--modified 
and of the standard Wilson--Dirac operators respectively.
Note that the projection on the momentum $\vec{\theta}/L$
of one of the quark legs in equation~(\ref{eq:fpcontractions}) it is not
realized by summing on the lattice points with an exponential factor but
it is encoded in the $\theta$--dependence of the modified Wilson--Dirac
operator and, consequently, of its inverse $S(\theta;x,y)$. 
This correlation is expected to decay exponentially at large times
as
\beq
f_P^{ij}(\theta;x_0) \qquad \stackrel{x_0\gg 1}{\longrightarrow} \qquad f_{ij}\  e^{- ax_0 E_{ij}(\theta,a)}
\eeq
where, a part from corrections proportional to the square of the lattice spacing, $E_{ij}$ is the
physical energy of the mesonic state.
After the continuum extrapolations one has to recover the expected relativistic
dispersion relations
\beq
E_{ij}^2 = M_{ij}^2+\left(\frac{\vec{\theta}}{L}\right)^2
\label{eq:drcont}
\eeq
where $M_{ij}$ is the mass of the pseudoscalar meson made of a $i$ and a $j$
quark anti--quark pair.

All the numerical results are obtained in the quenched
approximation of the QCD.
We have done simulations on a physical volume 
of topology $T\times L^3$ with $T=2L$ and linear extension~$L=3.2\ r_0$,
where $r_0$ is defined in~\cite{Sommer:1994ce}. 
All the missing parameters
of the simulations are given in table~1 of ref.~\cite{deDivitiis:2004kq}.

Setting the lattice scale by using the physical value $r_0= 0.5$ fm, the expected values of the physical momenta
associated with our choices of $\vec{\theta}$ are calculated according
to the following relation
\beq
|\vec{p}| = 
 \frac{|\vec{\theta}|}{L} 
 \simeq 0.125\ |\vec{\theta}| \; \mbox{GeV}
= \left\{
\begin{array}{l}
0.000 \\
0.217 \\
0.433 \\
0.650 \\
\end{array}  \right. \; \mbox{GeV} 
\eeq
where $L \simeq 1.6$ fm.
These values have to be compared with the value of the lowest physical momentum allowed
on this finite volume in the case of periodic boundary conditions, 
i.e. $|\vec{p}| \simeq 0.785$ GeV.
\begin{figure}[t]
\begin{center}
\includegraphics[width=\columnwidth]{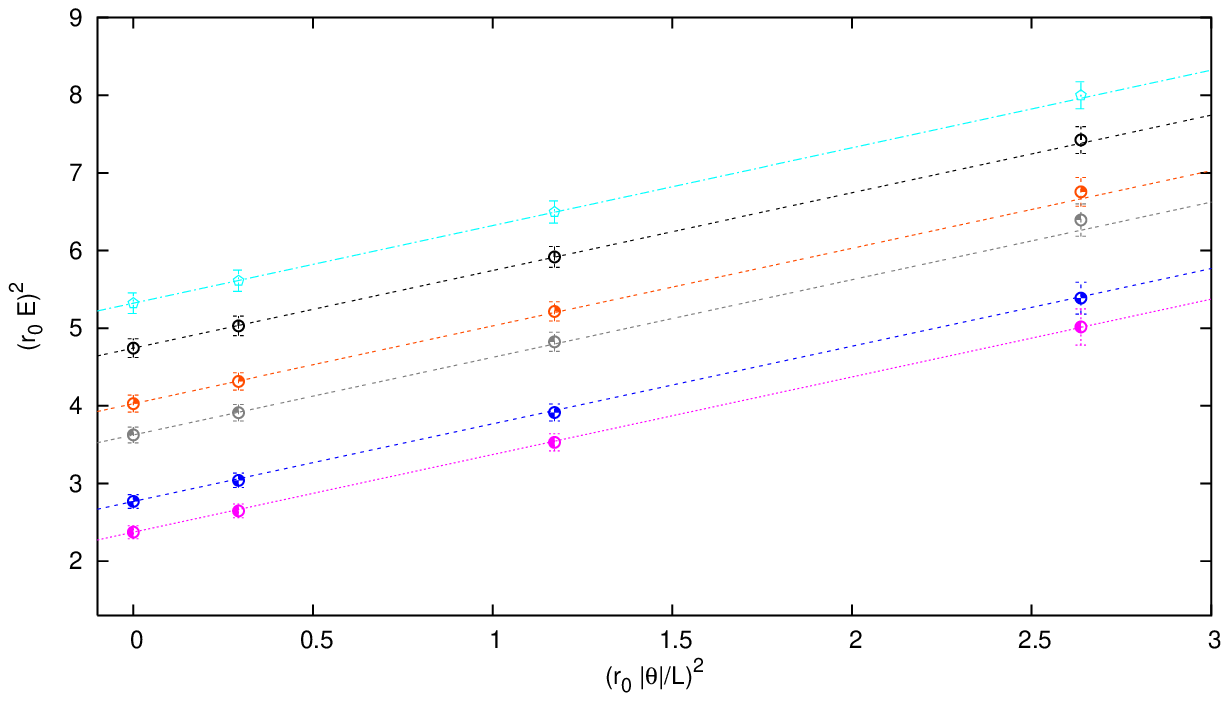}
\vskip -20pt
\mycaption{
Continuum dispersion relations. The data correspond to different
combinations of the simulated quark masses and reproduce very well
the expected theoretical behavior, i.e. straight lines having
as intercepts the meson masses and as angular coefficients one (see eq.~\ref{eq:drcont}).
}
\label{fig:disprel}
\end{center}
\vskip -20pt
\end{figure}

At fixed cut--off, for each combination of flavor indices and for each value of $\vec{\theta}$ 
we have extracted the 
effective energy from the correlations of eq.~(\ref{eq:fp}), $f_P^{ij}(\theta;x_0)$.
In fig.~\ref{fig:effmass} we show this quantity for the simulation performed at $\beta = 6.211$
corresponding to $r_0\ m_1^{RGI} = 0.655$ and $r_0\ m_2^{RGI} = 0.354$, for each simulated value
of $\vec{\theta}$. As can be seen the correlations with higher values of $|\vec{\theta}|$ are
always greater than the corresponding ones with lower values of the physical momentum,
a feature that will be confirmed in the continuum limit.
Being interested in the ground
state contribution to the correlation of eq.~(\ref{eq:fp}), we have averaged the effective energies
over a ground state plateau of physical length depending upon the quark flavors and we
have extrapolated the resulting quantities, $E^{ij}(\theta,a)$, 
to the continuum (see ref.~\cite{deDivitiis:2004kq} for details). 

The continuum results verify very well the dispersion relations of
equation~(\ref{eq:drcont}) as can be clearly seen from fig.~\ref{fig:disprel} in which
the square of $E^{ij}(\theta)$ for various combinations of the flavor indices is plotted
versus the square of the physical momenta $|\vec{\theta}|/L$. 
The plotted lines have not been fitted but have been obtained by using as
intercepts the simulated meson masses and by fixing their 
angular coefficients to one.



\end{document}